\documentclass[aps,pre,twocolumn,showpacs,epsfig,floats]{revtex4-1}
\usepackage{amsmath}
\usepackage{color}
\usepackage{graphicx}
\usepackage{epsfig}
\usepackage{amsfonts}
\usepackage{mathrsfs}
\usepackage{amsmath}

\begin{document}

\title{Quantum Signature of Chaos and Thermalization in Kicked Dicke Model}
 
\author{S. Ray, A. Ghosh and S. Sinha}
\affiliation{Indian Institute of Science Education and
Research-Kolkata, Mohanpur, Nadia-741246, India}
 
\date{\today}

\begin{abstract}

We study the quantum dynamics of the kicked Dicke model(KDM) in terms of the Floquet operator and analyze the connection between the chaos and thermalization in this context. The Hamiltonian map is constructed by taking the classical limit of the Heisenberg equation of motion suitably to study the corresponding phase space dynamics which shows a crossover from regular to chaotic motion by tuning the kicking strength. 
The fixed point analysis and calculation of the Lyapunov exponent(LE) provides us a complete picture of the onset of chaos in phase space dynamics. We carry out the spectral analysis of the Floquet operator which include the calculation of quasienergy spacing distribution, structural entropy and show the correspondence to the random matrix theory in the chaotic regime. Finally, we analyze the thermodynamics and statistical properties of the bosonic sector as well as the spin sector and discuss how such periodically kicked system relaxes to a thermalized state in accordance with the laws of statistical mechanics. We introduce the notion of an effective temperature and show that a microcanonical picture is emerging out in the thermodynamic limit indicating the thermalization occurring in such system.   

\end{abstract}


\maketitle

\section{Introduction}

The emergence of chaos in classical dynamical systems, in the long time limit, 
can often be related to the ergodic properties of the phase space \cite{Ruelle},
however, the process of equilibration of an isolated  quantum system is not fully understood.
The progress in quantifying the signature of chaos in quantum dynamics \cite{Haake_b} 
have generated interests towards understanding thermalization in isolated quantum systems \cite{MRigol_rev,Santos_rev}.
In an important foray, it had been shown that statistical properties of a closed quantum system under non-equilibrium evolution can be explained by eigenstate thermalization hypothesis(ETH) which proposes that thermalization happens at the level of individual many-body eigenstates \cite{Srednicki,Rigol_Nat}.
In a number of theoretical works it has been shown that strongly interacting many-body systems thermalize \cite{Rigol_Nat,Rigol_interacting} typically by energy sharing between many interacting modes \cite{MOlshanii,Izrailev}. 
An alternative viewpoint has been proposed in understanding thermalization of two-level interacting systems, namely the `Dicke model'(DM) \cite{Dicke}, by analyzing the semiclassical dynamics of a classically chaotic counterpart \cite{Altland,Feshke}.
This work indicates that the low-dimensional chaotic dynamics is the main criteria and route to thermalization in contrast to the complexity of the many body systems.
This also opens the possibility to study thermalization in relatively simple kicked systems with few degrees of freedom exhibiting chaotic dynamics \cite{Haake_b,Stockman,Casati}.
It is thus important to investigate the fate of thermalization in a simple entangled system in the presence of kicks.
Furthermore we ask the question whether it is possible to predict equilibration in such kicked systems from the underlying chaotic classical dynamics. 
To answer these questions, in this work we consider the Dicke model subjected to periodic 
kicks and demonstrate that this simple generalization induces
rich dynamical features in the classical as well as in the quantum regimes
leading to the process of thermalization.

Originally, Dicke model was proposed to describe an ensemble of two-level atoms interacting with a single mode of electromagnetic field, which shows a quantum phase transition from the normal phase of the atoms to the superradiant phase at a critical coupling strength \cite{Dicke,Brandes}. 
In recent experiments on ultracold atoms, Dicke like models have also been realized by coupling a Bose-Einstein condensate with the cavity mode \cite{Dicke_exp,esslinger_rmp}. These experiments also generated a new impetus to study non-equilibrium dynamics and quantum phases of Dicke like models \cite{Dicke_neq,Bastidas,domokos,Simons,Solenov,Sinha}. 
In another set of recent experiments on a system of ultra-cold atoms \cite{Raizen_kicked,Jessen_kicked} 
have further renewed interests in studying many-body dynamics in the presence of kicking \cite{DSen,Polkovnikov_kicked,Basu} 
and thermalization induced by periodic drive \cite{Polkovnikov_drive,Rigol_drive,Nandkishore}.
The above experiments motivate us to consider the Dicke model subjected to temporal perturbation by introducing a time periodic impulsive interaction between the spin and the bosonic mode. 
Thus by proposing the `kicked Dicked model' the aim of the present work is twofold: 
i) to present a detailed comparative analysis of classical and quantum dynamics of KDM;
ii) to study the spectral statistics of such kicked quantum system and elucidate the dynamical route to thermalization. 

The rest of the paper is organized as follows. In Sec.\ \ref{Model} we introduce the Hamiltonian of the KDM and obtain the Heisenberg equation of motion for the corresponding observables. This is followed by Sec.\ \ref{Class_Dyn} where we analyze the classical phase space dynamics by constructing a classical Hamiltonian map. Next, in Sec.\ \ref{Quan_Dyn} we discuss the quantum dynamics of KDM in terms of the Floquet operator and its spectral statistics. This analysis is carried out in two subsections : in subsection.\ \ref{Class-Quan}, a comparison between  classical and quantum dynamics is presented.
This is followed by subsection.\ \ref{Lev_stat} where we discuss the signature of chaos in terms of the spectral statistics of the Floquet operator and show the correspondence to the random matrix theory for large strength of kicking. 
The fate of thermalization in such periodically kicked system and the
emergence of statistical distribution in both the bosonic bath and the spin sub-system are presented in Sec.\ \ref{Therm}.
Finally in Sec.\ \ref{Conclu}, we summarize our results, discuss possible experiments relevant to our theory, and conclude.          

\section{The Model}
\label{Model}

The Dicke Model describes an ensemble of the two level atoms with an energy gap $\hbar \omega_0$ represented by the spin-1/2 objects interacting with a single cavity mode. Hence the collection of such `N' atoms can be represented by $\hat{S}_i=\sum_{j=1}^N \hat{\sigma}_i^{(j)}/2$ where $\hat{S}_i(i=x,y,z)$ represents the components of the total spin S. Thus at zero temperature DM represents a large spin of magnitude $S=N/2$ interacting with a single bosonic mode described by the Hamiltonian, $\hat{H} = \hat{H}_0 + \hat{H}_c$ \cite{Dicke,Brandes} with

\begin{subequations}
\begin{eqnarray}
\hat{H}_0 &=& \omega _0 \hat{S}_z + \Omega \hat{a}^{\dagger}\hat{a} \label{H_0} \\
\hat{H}_c &=& \frac{\lambda}{\sqrt{S}}(\hat{a}^{\dagger} + \hat{a})\hat{S}_x
\label{H_c}
\end{eqnarray} 
\label{Dicke_ham}
\end{subequations}
where $\hat{a}(\hat{a}^{\dagger})$ are the annihilation(creation) operators of the bosonic mode of frequency $\Omega$ and $\hat{H}_c$ represents the coupling term with the coupling constant $\lambda$. The single bosonic mode can be alternatively viewed as a quantum harmonic oscillator with mass `m' and the canonically conjugate position and momenta are given by $\hat{Q}_H = \sqrt{\frac{1}{2m\Omega}}(\hat{a}^{\dagger} + \hat{a})$ and $\hat{P}_H = i\sqrt{\frac{m\Omega}{2}}(\hat{a}^{\dagger} - \hat{a})$ respectively.   
Here and in the rest of the paper we set $\hbar = 1$, measure all energy(time) in units of $\omega_0(1/\omega_0)$ e.g. $\Omega \equiv \Omega/\omega_0$ and introduce the dimensionless variables of the oscillator(bosonic mode) $\hat{Q} \equiv \sqrt{m\Omega}\hat{Q}_H$ and $\hat{P} \equiv \hat{P}_H/\sqrt{m \Omega}$.

As a time-dependent generalization of the Dicke model we introduce a periodic kicking to the interaction between the spin and the bosonic mode. Thus the Hamiltonian of the kicked Dicke model is described by, 

\begin{subequations}
\begin{eqnarray}
\hat{H}(t) &=& \hat{H}_0 + \hat{H}_c(t), \label{Dicke_t} \\
\hat{H}_c(t) &=& \frac{\lambda _0}{\sqrt{S}} (\hat{a}^{\dagger} + \hat{a})\hat{S}_x \sum _{n=-\infty}^{\infty}\delta (t-nT), \label{Dicke_Hct}
\end{eqnarray}
\label{Dicke_hamt}
\end{subequations}

where $\lambda_0$ represents the kicking strength(equivalently the coupling strength) and T is the time period between the two consecutive kicking. The generic unitary time evolution of a system under the time-dependent Hamiltonian $\hat{H}(t)$ can be described by, $\vert \Psi(t) \rangle = \hat{\mathcal{T}} e^{-i \int _{0}^{t} \hat{H}(t) dt} \vert \Psi(0) \rangle$, where $\hat{\mathcal{T}}$ is the time ordering operator and $\Psi(t)$ is the wavefunction at time t. We note that, in the KDM the system evolves freely in between the successive kicks under $\hat{H}_0$; this free time evolution is governed by the unitary operator $\hat{U}_0 = e^{-i \hat{H}_0 T}$. This is followed by the instantaneous kick to the system which can be described by the unitary operator $\hat{U}_c = e^{-i\frac{\lambda _0}{\sqrt{S}}(\hat{a}^{\dagger} + \hat{a})\hat{S}_x}$. So the total time evolution of the system within a complete cycle under $\hat{H}(t)$\ref{Dicke_hamt} is described by the Floquet operator \cite{Haake_b,Stockman} which is given by, 

\begin{equation}
\hat{\mathcal{F}} = \hat{U}_c \hat{U}_0
\label{Floquet_op}
\end{equation}

Having defined the Floquet operator, it is now easy to describe the discrete time evolution of any operator $\hat{O}$ in the Heisenberg picture; the corresponding equation of motion can be written as $\hat{O}^{(n+1)} = \hat{\mathcal{F}}^{\dagger} \hat{O}^{(n)} \hat{\mathcal{F}}$, where $\hat{O}^{(n)}$ is the operator at the time t = nT. Following such procedure, we obtain the stroboscopic time evolution of the operators corresponding to the spin components and the bosonic mode(harmonic oscillator) given by,

\begin{subequations}
\begin{eqnarray}
\hat{S}_{x}^{(n+1)} &=& \hat{S}_{x}^{(n)}\cos T - \hat{S}_{y}^{(n)}\sin T \label{Sx} \\
\hat{S}_{y}^{(n+1)} &=& \cos \hat{z} (\hat{S}_{x}^{(n)}\sin T + \hat{S}_{y}^{(n)}\cos T) - \hat{S}_{z}^{(n)}\sin \hat{z} \label{Sy} \\
\hat{S}_{z}^{(n+1)} &=& \sin \hat{z} (\hat{S}_{x}^{(n)}\sin T + \hat{S}_{y}^{(n)}\cos T) + \hat{S}_{z}^{(n)}\cos \hat{z} \label{Sz} \\
\hat{Q}^{(n+1)} &=& \hat{Q}^{(n)}\cos \Omega T + \hat{P}^{(n)}\sin \Omega T \label{Q} \\
\hat{P}^{(n+1)} &=& \hat{P}^{(n)}\cos \Omega T - \hat{Q}^{(n)}\sin \Omega T - \lambda _0 \sqrt{\frac{2}{S}}\left[\hat{S}_{x}^{(n)}\cos T 
\right.\nonumber\\ 
&&\left. - \hat{S}_{y}^{(n)}\sin T \right] \label{P} \\
\text{where} &\quad & \hat{z} = \lambda _0\sqrt{\frac{2}{S}}\left[\hat{Q}^{(n)}\cos \Omega T + \hat{P}^{(n)}\sin \Omega T \right]
\end{eqnarray}
\label{Heisen_eqn}
\end{subequations}

\section{Classical Dynamics and Fixed Point Analysis}
\label{Class_Dyn}

In this section we analyze the classical counterpart of the Kicked Dicke Model by taking the classical limit of the corresponding Heisenberg equations of motion obtained in Eq.\ \ref{Heisen_eqn}. 
The classical limit can be achieved for the large spin i.e. $S \rightarrow \infty$.
It can be noted that the spin variables$(\hat{S}_{x,y,z})$ can be classically described by the components of a spin vector $\vec{S} \equiv (S\sin\theta \cos\phi, S\sin\theta \sin\phi, S\cos\theta)$, where $\theta$ and $\phi$ are the polar and the azimuthal angles respectively, representing the orientation of the spin vector.
We rescale the classical spin variables by S i.e. $s_i \equiv \hat{S}_i/S$; the conservation of the total spin leads to the constraint $s_x^2 + s_y^2 + s_z^2 = 1$. In addition to rescaling the spin variables, we appropriately scale the canonically conjugate variables of the oscillator $\hat{P}$ and $\hat{Q}$ by $\sqrt{S}$, i.e., $q \equiv \hat{Q}/\sqrt{S}$ and $p \equiv \hat{P}/\sqrt{S}$. With these transformations, we note that $\left[s_i,s_j\right]=i\epsilon _{ijk}s_k/S$ and $\left[q,p\right]=i/S$; hence in the limit $S \rightarrow \infty$ it is easy to see that the commutators vanish as the variables become classical. From Eq.\ \ref{Heisen_eqn}, the stroboscopic time evolution of the corresponding classical dynamical variables in between the consecutive kicks are described by the following Hamiltonian map,

\begin{subequations}
\begin{eqnarray}
s_{x}^{(n+1)} &=& s_{x}^{(n)}\cos T - s_{y}^{(n)}\sin T \label{Sx_c} \\
s_{y}^{(n+1)} &=& \cos z(s_{x}^{(n)}\sin T + s_{y}^{(n)}\cos T) - s_{z}^{(n)}\sin z \label{Sy_c} \\
s_{z}^{(n+1)} &=& \sin z(s_{x}^{(n)}\sin T + s_{y}^{(n)}\cos T) + s_{z}^{(n)}\cos z \label{Sz_c} \\
q^{(n+1)} &=& q^{(n)}\cos \Omega T + p^{(n)}\sin \Omega T \label{Q_c} \\
p^{(n+1)} &=& -q^{(n)}\sin \Omega T + p^{(n)}\cos \Omega T - \sqrt{2} \lambda _0 \left[s_{x}^{(n)}\cos T \right.\nonumber\\
&&\left. - s_{y}^{(n)}\sin T\right] \label{P_c} \\
\text{where} &\quad & z = \sqrt{2} \lambda _0 \left(q^{(n)}\cos \Omega T + p^{(n)}\sin \Omega T \right)
\end{eqnarray}
\label{Ham_map}
\end{subequations} 

\begin{figure}[ht]
\begin{center}
\includegraphics[scale=0.114]{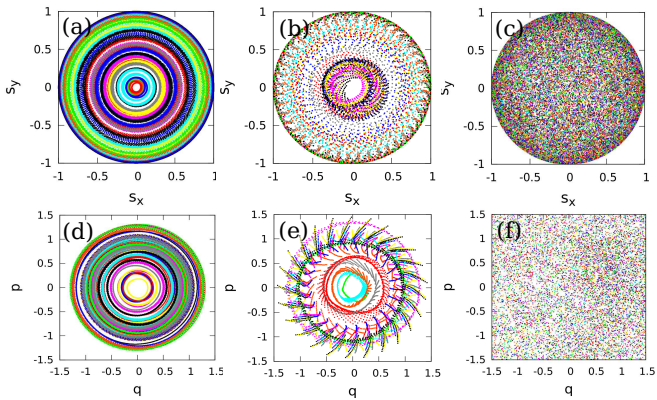}
\end{center}
\caption{Phase portrait obtained by iterating the Hamiltonian map in Eq.\ \ref{Ham_map} has been shown for increasing kicking strength $\lambda _0$. Top panel represents the phase portraits in the projected spin space($s_x$-$s_y$ plane) and bottom panel represents the phase portraits in the bosonic space($q$-$p$ plane). Columns from left to right correspond to $\lambda _0 = 0.01, 0.05$ and $0.6$ respectively. We have set $\Omega$ = 0.5, T = 1.0.}
\label{Phase_Port}
\end{figure}

By iterating the Hamiltonian map described in Eq.\ \ref{Ham_map} we obtain the phase space dynamics of the bosonic sector as well as the spin dynamics projected in x-y plane of the Bloch sphere. We observe that for smaller values of the kicking strength $\lambda _0$ phase space trajectories are regular whereas the system crosses over to a chaotic phase space dynamics (for both spin and oscillator) with increasing $\lambda _0$ as depicted in Fig.\ \ref{Phase_Port}.

To quantify the onset of chaos we analyze the Hamiltonian map given in Eq.\ \ref{Ham_map}. The fixed points are obtained from the condition $s_{i}^{(n+1)} = s_{i}^{(n)} = s_i^*$, $q^{(n+1)} = q^{(n)} = q^*$ and $p^{(n+1)} = p^{(n)} = p^*$. We note the existence of the trivial fixed points given by $\{s_x^*, s_y^*, s_z^*, q^*, p^*\} \equiv \{0,0,\pm 1,0,0\} \equiv s^*_{\pm}$ corresponding to the excited state and ground state of the DM(Eq.\ \ref{Dicke_ham}) for all values of the kicking strength $\lambda_0$. Apart from that, the non-trivial fixed points can be obtained by solving the non-linear equation for $q^*$ which is given as follows,

\begin{equation}
\frac{(q^*)^2 \tan^2\left(\frac{\Omega T}{2}\right)}{\cos^2\left(\frac{T}{2}\right)} \left[1 + \cot^2\left(\frac{\lambda _0 q^*}{\sqrt{2}}\right) \sin^2\left(\frac{T}{2}\right) \right] = \lambda _0^2/2. 
\label{Q_c_sol}
\end{equation}

We note that Eq.\ \ref{Q_c_sol} admits only the non-zero solution of $q^*$. The non-zero values of the other variables corresponding to this fixed point are given by,

\begin{subequations}
\begin{eqnarray}
p^* &=& \tan \left(\frac{\Omega T}{2}\right) q^* \label{P_c_sol} \\
s_x^* &=& -(\sqrt{2}/\lambda _0)\tan\left(\frac{\Omega T}{2}\right)q^* \label{Sx_c_sol} \\
s_y^* &=& \tan\left(\frac{T}{2}\right)(\sqrt{2}/\lambda _0)\tan\left(\frac{\Omega T}{2}\right)q^* \label{Sy_c_sol} \\
s_z^* &=& \cot\left(\frac{\lambda _0 q^*}{\sqrt{2}}\right)\left[\sin(T)s_x^* + \cos(T)s_y^*\right] \label{Sz_c_sol}
\end{eqnarray}
\label{fixed_point}
\end{subequations}

\begin{figure}[ht]
\begin{center}
\includegraphics[scale=0.318]{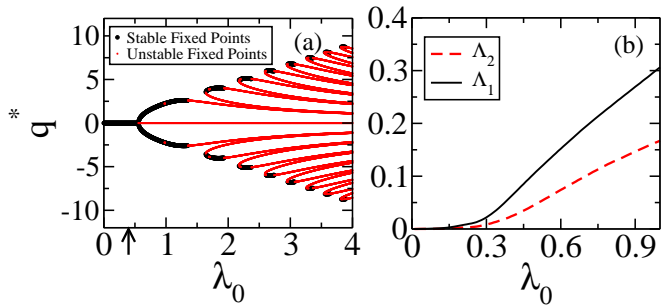}
\end{center}
\caption{(a) Fixed points $q^*$ has been shown as a function of $\lambda _0$. The bold points(black) indicate the stable fixed points and dots(red) indicate the unstable fixed points. The arrowhead indicates the critical kicking strength $\lambda_c^{\prime} \sim 0.39$ at which $s^*_{+}$ loses the stability. (b) The two positive Lyapunov exponents $\Lambda_1$ and $\Lambda_2$ have been shown as a function of $\lambda_0$. The other parameters we have set for this figure are $\Omega$ = 0.5, T = $\pi/3$.}
\label{Stability}
\end{figure}

From Eq.\ \ref{Q_c_sol} we note that the non-trivial fixed points always appear in pairs (with equal magnitude and opposite sign). By solving Eq.\ \ref{Q_c_sol} we notice that the first pair of non-trivial fixed points appear above the critical coupling strength $\lambda _c$ given by,

\begin{equation}
\lambda_c = \sqrt{2\tan(\Omega T/2)\tan(T/2)}.
\label{l_c}
\end{equation} 
Additional pairs of non-trivial fixed points appear at $\lambda _0 \approx \lambda _c\sqrt{\frac{(2n+1)\pi}{2\sin(T/2)}}$ where n is a positive integer denoting the no of pair as shown in Fig.\ \ref{Stability}(a). Next we analyze the linear stability of these fixed points; to do that we construct the Jacobian matrix J whose elements are given by, $J_{ij} = \partial A_i^{(n+1)}/\partial A_j^{(n)}$ where $\{A_1,A_2,A_3,A_4,A_5\} \equiv \{s_x, s_y, s_z, q, p\}$ and study the nature of the eigenvalues of J \cite{Strogatz}. We find that out of the five complex eigenvalues of J, one always comes out to be $1$ as a consequence of the fixed magnitude of the spin vector and the remaining four eigenvalues appear in pairs (complex conjugate of each other). Stability of the fixed points are ensured if the magnitudes of the eigenvalues are unity \cite{Haake_b,Strogatz}.
First we note that the two trivial fixed points $s^*_{+}$ and $s^*_{-}$ corresponding to the excited and ground state of the DM respectively remain dynamically stable for smaller kicking strength. With the increase in $\lambda_0$, $s^*_{+}$ loses the stability at the critical kicking strength $\lambda_c^{\prime}$ given by, 
\begin{equation}
\lambda_c^{\prime} = \sqrt{(\cos T - \cos \Omega T)^2/(2\sin T\sin \Omega T)}
\end{equation}
Further increase in $\lambda_0$ results in the instability of the fixed point $s^*_{-}$ corresponding to the ground state at the critical kicking strength $\lambda_c$(Eq.\ \ref{l_c}).
In Fig.\ \ref{Stability}(a) $q^*$ has been shown as a function of $\lambda _0$; we see that $s^*_{+}$ loses the stability at $\lambda_c^{\prime}$(shown in arrowhead) much before $\lambda_c$ at which the other trivial fixed point $s^*_{-}$ becomes unstable. For $\lambda_0 > \lambda_c$ the non-trivial fixed points obtained uniquely from non-zero $q^*$ becomes dynamically stable. We further notice that the region of stability of these non-trivial fixed points decrease with increasing coupling strength and finally the phase space becomes completely chaotic for large $\lambda_0$ with no such stable fixed point in the system.

In a dynamical system the `Lyapunov Exponent'(LE) is a measure to quantify the onset of chaos. 
In the present case the LE's of the Hamiltonian map(Eq.\ \ref{Ham_map}) are estimated by the method of QR decomposition of the corresponding linearized system \cite{Lauterborn_lyap}.
We represent the converged values of the Lyapunov exponents as (0,$\pm \Lambda _2$, $\pm \Lambda _1$) where $\Lambda_1 > \Lambda_2$. 
In Fig.\ \ref{Stability}(b) we have shown the two positive Lyapunov exponents $\Lambda _{1,2}$ (averaged over 1000 initial conditions) as a function of the coupling strength $\lambda _0$. We note that for small kicking strength $\Lambda_{1,2} \sim 0$; LEs start rising around the kicking strength $\lambda_c^{\prime}$ and increase with increasing $\lambda _0$ as depicted in Fig.\ \ref{Stability}(b).  
From the above analysis, it has become clear that the classical counterpart of the KDM undergoes a crossover from regular to chaotic phase space dynamics by tuning the kicking strength and the onset of chaos can be quantified in terms of the average LE.

\section{Quantum Dynamics}
\label{Quan_Dyn}

In this section we study the stroboscopic quantum dynamics of the kicked Dicke model in terms of the Floquet operator and discuss its spectral properties. 
Such an analysis is carried out considering the computational basis as $|\chi\rangle$ which is the simultaneous eigenstate of $\hat{S}_{z}$ and $\hat{a}^{\dagger}\hat{a}$.  
In the Schr\"odinger picture the state of the system at $t=nT$, being evolved under the kicked Dicke Hamiltonian(Eq.\ \ref{Dicke_hamt}), can be written as follows, 
\begin{equation}
\vert \psi(nT)\rangle = \hat{\mathcal{F}}^n \vert \psi(0)\rangle
\end{equation}
where $\hat{\mathcal{F}}$ is the Floquet operator (see Eq.\ \ref{Floquet_op}) and $\vert \psi(0)\rangle$ is the initial state.
Now the spectral decomposition of the wavefunction in the eigenbasis of $\hat{\mathcal{F}}$ allows to write down the time evolved state at $t=nT$ in the following way,
\begin{equation}
\vert \psi(nT) \rangle = \sum _{\nu} c_{\nu} e^{-i \phi_{\nu} n} \vert \Phi _{\nu} \rangle
\end{equation}
where the co-efficient $c_j = \langle \Phi _j \vert \psi(0) \rangle$, $e^{-i \phi_{\nu}}$ and $\vert \Phi_{\nu} \rangle$ are the eigenvalue and the eigenvector corresponding to the $\nu$th mode, $\phi_{\nu}$ being the eigenphase, of the Floquet operator $\hat{\mathcal{F}}$. At any time instant $t = nT$ the expectation value of any observable $\hat{O}$ can be obtained from $\langle \hat{O}  \rangle = \langle \psi(nT) \vert \hat{O} \vert \psi(nT) \rangle$.

Following such prescription, we aim to present the classical-quantum correspondence and discuss the fate of classicality in terms of the coherent state in different regimes of the coupling strength $\lambda _0$ in section\ \ref{Class-Quan}. Next, in section\ \ref{Lev_stat} we will analyze the spectral properties of the Floquet operator $\hat{\mathcal{F}}$ and discuss the correspondence to the phase space dynamics.  

\subsection{Classical-Quantum Correspondence}
\label{Class-Quan}

In this subsection we compare the classical dynamics obtained by iterating the Hamiltonian map (Eq.\ \ref{Ham_map}) with the stroboscopic quantum dynamics in the different regime of the kicking strength.
A suitable classical representation of the quantum state $\vert \psi \rangle$ is given by the coherent state
$\vert \alpha ; \theta, \phi \rangle = \vert \alpha \rangle \otimes \vert \theta, \phi \rangle$, where $\vert \alpha \rangle$ represents the coherent state of the harmonic oscillator \cite{sudarsan} given by 
\begin{equation}
\vert \alpha \rangle = e^{-\vert \alpha \vert ^2/2} e^{\alpha \hat{a}^{\dagger}}\vert 0 \rangle
\end{equation}
where $\alpha$ is the classical variable corresponding to the operator $\hat{a}$.
$\vert \theta, \phi \rangle$ is the spin coherent state \cite{Radcliff} given by,
\begin{equation}
\vert \theta, \phi \rangle = (1 + \vert z \vert ^2)^{-S} e^{z \hat{S}_{+}} \vert S,-S \rangle
\end{equation}
where $\theta$ and $\phi$ are polar and azimuthal angles respectively, representing the orientation of the spin variables of magnitude S and $z = e^{-i \phi} \tan \theta/2$. 

We choose the initial state $\vert \psi(0) \rangle$ as the coherent state 
$\vert \alpha ; \theta, \phi \rangle$ which is constructed from the values of the corresponding classical variables. Such a choice of the initial state is relevant to analyze the classical-quantum correspondence. 
\begin{figure}[ht]
\centering
\includegraphics[scale=0.31]{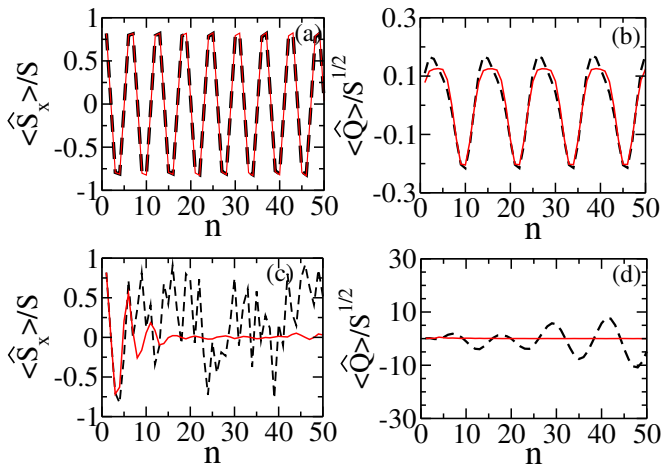}
\caption{Time evolution of $\langle \hat{S}_x \rangle/S$ and $\langle \hat{Q} \rangle/\sqrt{S}$ have been shown for $\lambda _0=0.05$ in (a,b) and for $\lambda _0=1.0$ in (c,d). Solid lines(red) correspond to the time evolution of the respective quantities obtained from quantum dynamics and dashed lines(black) represent the same obtained by iterating the classical dynamical map in Eq.\ \ref{Ham_map}. We have chosen $S = 20$ and $N_{max} = 30$ for the quantum dynamics. Other parameters chosen for this plot are $T=\pi/3$, $\Omega=0.5$ and the initial coherent state has been constructed using $q = 0.11$, $p = 0.14$, $\theta = 1.2$ and $\phi = 0.5$.}
\label{class_quan}
\end{figure}
First, we compute the average values of the relevant observables e.g. $\langle \hat{S}_x \rangle$, $\langle \hat{Q} \rangle$ which has been shown as a function of time in Fig.\ \ref{class_quan} for different coupling strengths. 
We see that for weak kicking strength, the time evolution of both the quantities are in good agreement with the corresponding classical variables obtained by iterating the Hamiltonian map. Whereas for large kicking strength, the results obtained from quantum dynamics deviates significantly from its classical counterpart. 
We further emphasize on the fluctuations around these mean values calculated in the regular regime (for small $\lambda _0$) and in the chaotic regime (for large $\lambda _0$). The root mean square deviation around the mean values of an observable $\hat{O}$ can be written as,
\begin{equation}
\delta O = \left(\langle \hat{O}^2 \rangle - \langle \hat{O} \rangle ^2 \right)^{1/2}
\end{equation} 
In Fig.\ \ref{coherence}(a,b) we have shown $\delta Q$ and $\delta P$ as a function of time for different kicking strength. We find that in the weak coupling regime, the fluctuations are very small, whereas these increase with increasing kicking strength.
This result originates from the fact that in the weak coupling regime the coherent nature of the wavefunction persists, whereas in the strong coupling regime it indicates the loss of the classicality in terms of the coherent state. 

\begin{figure}[ht]
\centering
\includegraphics[scale=0.33]{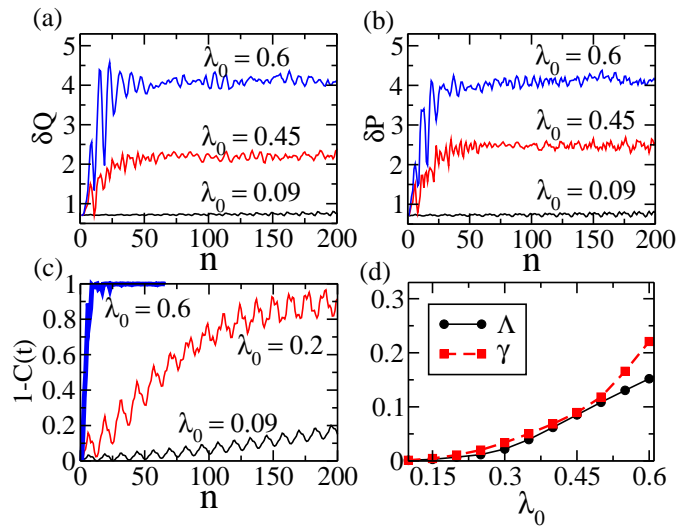}
\caption{(a)-(b) $\delta Q$ and $\delta P$ have been shown as a function of time. (c) The deviation of the time evolved wavefunction from the coherent state, characterized by 1-C(t) as a function of the time step(n) has been shown for different values of the coupling strength $\lambda _0$. (d) Rate of loss of coherence $(\gamma)$ and its resemblance with the highest LE $(\Lambda)$ has been shown as a function of $\lambda_0$. The other parameters are same as in Fig.\ \ref{class_quan}.}
\label{coherence}
\end{figure}
To further elucidate this fact we compute `coherence factor' which gives the estimate of the overlap of the time-evolved wavefunction with the corresponding coherent state and is defined in the following way,
\begin{equation}
C(t) = \vert \langle \alpha; \theta, \phi \vert \psi(t) \rangle \vert ^2
\end{equation}  
where we construct the coherent state $\vert \alpha ; \theta, \phi \rangle$ at each instant of time where $\alpha$, $\theta$ and $\phi$ are computed from $\langle \hat{Q} \rangle$, $\langle \hat{P} \rangle$ and $\langle \hat{S}_{x,y,z} \rangle$. In Fig.\ \ref{coherence}(c) we have shown the quantity 1-C(t) as a function of time t to quantify the deviation of $\vert \psi(nT) \rangle$ from the coherent state at that instant of time. We see that for smaller value of $\lambda _0$, $1-C(t)\sim 0$ in the course of time evolution which is consistent with the fact that the two subspaces of the system remains non-entangled in the weak coupling regime and the semiclassical dynamics can be a good prescription to capture the dynamical features \cite{Altland,Feshke}. On the other hand in the chaotic regime, 1-C(t) grows faster with time and eventually saturates at $~1.0$. Furthermore, the growth rate of 1-C(t) or equivalently the rate of loss of coherence can be quantified by the exponential fitting parameter $\gamma$ which is surprisingly in strong resemblance with the Lyapunov exponent obtained from the classical dynamics and increases with $\lambda _0$ (see Fig.\ \ref{coherence}d). 
This observation indicates the fact that in the strong coupling regime when the corresponding phase space is chaotic, the coherent nature of the wavefunction is completely lost and this results in the loss of classicality of the system in terms of the coherent state prescription.

\subsection{Spectral Statistics}
\label{Lev_stat}
Two broad class of dynamics can also be characterized by the spectral statistics of the corresponding quantum Hamiltonian. As a consequence of Berry-Tabor conjecture, Poisson distribution of the energy level spacing implies regular phase-space dynamics\cite{Berry}. 
On the other hand, according to BGS conjecture \cite{Bohigas} Wigner-Dyson distribution of energy level spacing is an indicator of the onset of chaos in the classical dynamics.
In a periodically driven quantum system the onset of chaos can alternatively be captured by studying the level statistics of the quasi-energies of the corresponding Floquet operator\cite{Haake_b,Haake_p}.
We recall that due to unitarity, the eigenvalues of the Floquet operator $\hat{\mathcal{F}}$ can be written in the form of $e^{-i \phi_{\nu}}$, where $\phi_{\nu}$ is called `eigenphase' corresponding to the $\nu$th eigenvalue of $\hat{\mathcal{F}}$.
In this subsection we focus on the statistical properties of the eigenphases and eigenvectors of the Floquet operator$(\hat{\mathcal{F}})$. 

We note that the Dicke Hamiltonian in Eq.\ \ref{Dicke_ham} has a conserved parity corresponding to the parity operator $\hat{\Pi} = exp(i \pi \hat{N})$ \cite{Brandes}, where $\hat{N} = \hat{a}^{\dagger}\hat{a} + \hat{S}_z + S$. The parity operator possess two eigenvalues $\mathcal{P} = \pm 1$ corresponding to the odd or even eigenvalues of $\hat{N}$. We also observed that the parity is conserved in the Floquet dynamics. 
We separate out the eigenmodes of $\hat{U}$ based on the eigenvalues($\mathcal{P}$) of the parity operator $\hat{\Pi}$ and call the subspace to be even(odd) according as $\mathcal{P}=1(-1)$. The statistical analysis is performed separately for two symmetry sectors.

\begin{figure}[ht]
\begin{center}
\includegraphics[scale=0.32]{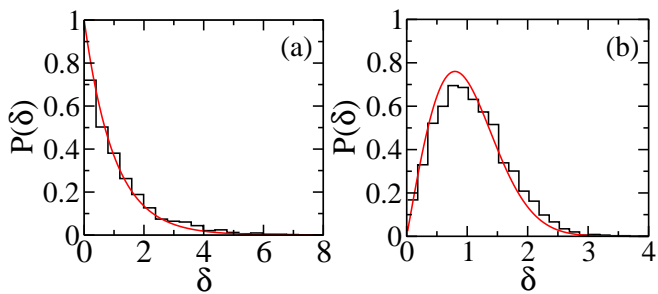}
\end{center}
\caption{Spacing distribution of the eigenphases are shown for $\lambda _0 (=0.1)$ in (a) and for $\lambda _0 (=10.0)$ in (b) considering the even subspace of $\hat{\mathcal{F}}$. For smaller kicking strength the distribution follows Poisson statistics, whereas for higher $\lambda_0$ the distribution follows Wigner-Surmise; the corresponding probability distributions are shown in solid curve(red). We have set $\Omega = 0.5$, $T = \pi/3$, $S = 25$ and truncate the bosonic Hilbert space at $N_{max} = 80$ for this figure.}
\label{lev_spacing}
\end{figure}

We numerically diagonalize the Floquet operator to obtain the eigenphases $\phi_{\nu}$ corresponding to a fixed symmetry sector and sort them in ascending order and calculate the level spacing $\delta_{\nu} = \phi_{\nu + 1} - \phi_{\nu}$.
The normalized spacing distribution of the eigenphases$(\phi_{\nu})$ is shown in Fig.\ \ref{lev_spacing}.  
We note that for small value of the coupling strength $\lambda _0$ the distribution follows the Poisson statistics given by $P(\delta) = \exp(-\delta)$, 
whereas in the limit of strong coupling, this distribution is in agreement with the prediction of orthogonal class RMT which can be effectively described by the Wigner-Surmise given by $P(\delta) = \frac{\pi \delta}{2}\exp\left(-\frac{\pi \delta ^2}{4}\right)$  \cite{Haake_b,Mehta} exhibiting the fact of level repulsion in the chaotic regime. 

\begin{figure}[ht]
\begin{center}
\includegraphics[scale=0.3]{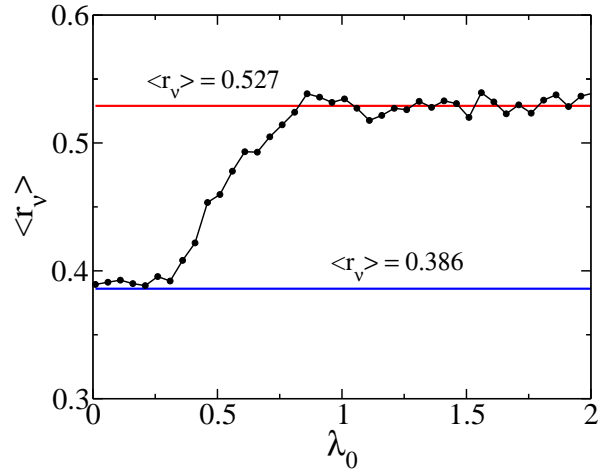}
\end{center}
\caption{Average value of the ratio $r_{\nu}$ of the level spacing is shown as a function of $\lambda _0$ for S = 30, $N_{max}$ = 60. The other parameters chosen for this plot are same as in Fig.\ \ref{lev_spacing}.}
\label{r_avrg}
\end{figure}

To study the crossover from Poisson to the Wigner-Dyson statistics of spacing distribution of the eigenphases we further calculate a dimensionless quantity $r_{\nu}$, which is defined as \cite{Huse,Bogomolny},
\begin{equation}
r_{\nu} = \frac{\text{min}(\delta _{\nu+1},\delta _{\nu})}{\text{max}(\delta _{\nu+1},\delta _{\nu})}
\end{equation}
where $\delta _{\nu}$ is the gap between consecutive eigenphases. For Poisson distribution of level spacing, it can be shown that $\langle r_{\nu} \rangle = 2 \ln 2 - 1 \approx 0.386$, whereas in case of circular orthogonal ensemble of RMT, $\langle r_{\nu} \rangle \approx 0.527$ \cite{Rigol_drive}. In Fig.\ \ref{r_avrg}, the variation of $\langle r_{\nu} \rangle$ with the licking strength $\lambda _0$ is depicted, which clearly indicates the crossover from Poisson to the Wigner-Dyson statistics resulting from level repulsion with increasing $\lambda _0$. 

More evidence of chaos can also be obtained from the study of eigenvectors of the Floquet operator $\hat{U}$. We decompose the eigenvectors $|\Phi_{\nu}\rangle$ of $\hat{U}$ in terms of the computational basis $|\chi\rangle$ of corresponding symmetry sector,
\begin{equation}
|\Phi_{\nu}\rangle = \sum_{\chi} c_{\chi}^{\nu} |\chi\rangle,
\end{equation}
where $|\chi\rangle$ is simultaneous eigenstate of $\hat{S}_{z}$ and $\hat{a}^{\dagger}\hat{a}$. 

\begin{figure}[ht]
\begin{center}
\includegraphics[scale=0.168]{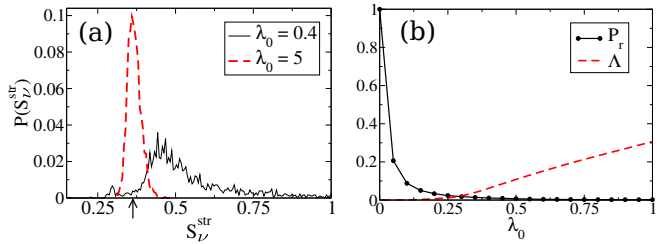}
\end{center}
\caption{(a) Distribution of the structural entropy $S^{str}_{\nu}$ of different eigenmodes($\nu$) of $\hat{\mathcal{F}}$ is shown for $\lambda _0 = 0.4$ (black solid line) and $\lambda _0 = 5.0$ (red dotted line) for $S=25$ and $N_{max}=70$. In the chaotic regime(for large $\lambda_0$) the distribution is peaked around $\sim 0.3646$ (pointed out by the arrowhead) which is consistent with the orthogonal ensemble of RMT. (b) Average participation ratio $P_r$ has been shown as a function of $\lambda_0$ exhibiting a strong resemblance with the highest Lyapunov exponent $\Lambda$.}
\label{str_entr}
\end{figure}

The degree of delocalization of the eigenmodes can be quantified by calculating the inverse participation ratio(IPR),
\begin{equation}
\quad \xi_{\nu} = \frac{1}{\sum_{\chi}\vert c_{\chi}^{\nu} \vert ^4}.
\end{equation}
To capture the crossover phenomena we calculate the average participation ratio which is given by $P_r=\sum_{\nu}P_{\nu}/\sum_{\nu}\nu$ where $P_{\nu}=1/\xi_{\nu}$ and the sum runs over the number of eigenmodes in the even(odd) subspace. 
From Fig.\ \ref{str_entr}(b), we see that with increasing $\lambda _0$, the participation ratio $P_r$ decreases and finally it vanishes indicating delocalization in chaotic regime which is evident from the growth of the Lyapunov exponent $\Lambda$ obtained from the classical dynamics. 
We note that all the eigenmodes of the Floquet operator equally participate in the classically chaotic regime.

To analyze the degree of mixing of the Floquet eigenvectors and its correspondence to RMT, we compute the structural entropy which is defined as \cite{Verga},
\begin{equation}
S_{\nu}^{str} = -\sum_{\chi}|c_{\chi}^{\nu}|^2 \ln |c_{\chi}^{\nu}|^2 - \ln \xi_{\nu},
\end{equation}
where $S_{\nu}$ is the structural entropy corresponding to the $\nu$th eigenmode.
It can be shown that for eigenvectors of the random matrices of orthogonal class, $S^{str}$ approaches to an universal value $\approx 0.3646$ irrespective of the dimension \cite{Santos_rev}. In Fig.\ \ref{str_entr}(a) we have shown the distribution of $S_{\nu}^{str}$ for small and large values of $\lambda _0$, which clearly shows that the distribution becomes sharply peaked around the value $0.3646$ for large $\lambda_{0}$.
The results presented in this section establish the connection between dynamical chaos in the kicked DM and the spectral statistics of the corresponding Floquet operator which exhibits its correspondence to RMT of orthogonal class for large kicking strength.

\section{Chaos and Thermalization}
\label{Therm}
The analysis in previous sections reveals that the kicked DM undergoes a crossover from regular motion to chaotic dynamics by tuning the kicking strength $\lambda_{0}$.
To this end we investigate how the spin sub-system and the bosonic bath thermalize as a consequence of chaos.

\begin{figure}[ht]
\centering
\includegraphics[scale=0.32]{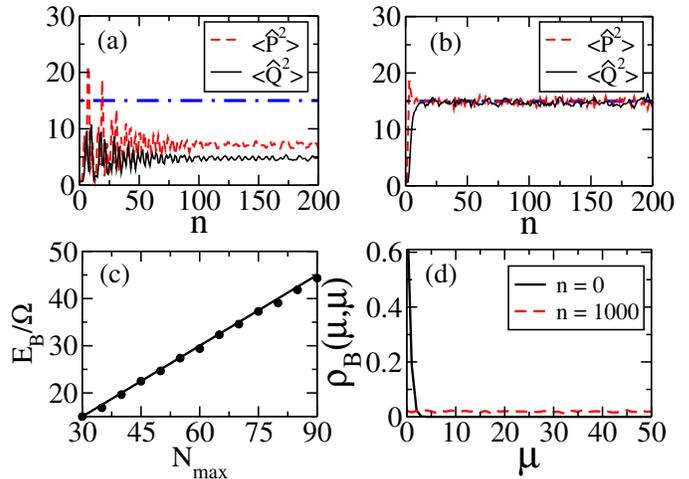}
\caption{$\langle \hat{P}^2 \rangle$ and $\langle \hat{Q}^2 \rangle$ as a function of the time step n is shown for (a) $\lambda _0 = 0.4$ and (b) $\lambda _0 = 3.0$ for $S=15$ and $N_{max}=30$. The dashed-dot lines(blue) in (a) and (b) denote the energy of the oscillator obtained from microcanonical ensemble approach i.e. $E_{B}/\Omega = N_{max}/2$. The linear growth of $E_{B}/\Omega$ with $N_{max}$ has been shown in (c); filled circles denote the saturated value of the average energy of the bosonic space($E_{B}/\Omega$) after long time evolution and the solid straight line represents the energy of the oscillator obtained from microcanonical ensemble i.e. $E_{B}/\Omega = N_{max}/2$. (d) Distribution of the diagonal elements of $\rho_B$ has been shown. Other parameters chosen are $T=\pi/3$ and $\Omega = 0.5$.}
\label{equip}
\end{figure}

First, we focus on the bosonic bath and calculate the thermodynamic quantities related to it. In Fig.\ \ref{coherence}(a) we have seen that for large $\lambda_0$ when the corresponding phase space is chaotic, although $\langle \hat{Q} \rangle,\langle \hat{P} \rangle \approx 0$, but the fluctuations $\delta Q$ and $\delta P$ take non-vanishing values which indicates a diffusive motion of the oscillator covering the full phase space.
We study the partitioning of energy between the degrees of freedom $P$ and $Q$ of the
oscillator(bosonic mode).
As shown in Fig.\ \ref{equip}(b), the steady state fluctuations satisfy the relation 
$\langle \hat{Q}^2 \rangle \simeq \langle \hat{P}^2 \rangle$, indicating the {\it equipartition of energy} in the chaotic regime; whereas for small $\lambda _0$ the respective quantities saturate to different values (see Fig.\ \ref{equip}(a)).
Although a preliminary signature of thermalization in chaotic regime is observed, more quantitative information can be obtained from the density matrices of the corresponding subsystems. The reduced density matrix $\hat{\rho}_{B(S)}$ of the boson(spin) at time $t=nT$ can be calculated from the relation,
\begin{equation}
\hat{\rho}_{B(S)} = Tr_{S(B)} \left( |\psi(nT)\rangle \langle \psi(nT)| \right),
\label{den_mat}
\end{equation}
where $|\psi(nT)\rangle$ represents the dynamical state of the total system and $Tr_{S(B)}(.)$ denotes partial trace with respect to spin(bosonic) degrees of freedom.
In the regime of strong coupling, we notice that starting from an arbitrary initial state the reduced density matrices evolve into a diagonal form in the basis of corresponding Hamiltonians of the subsystems.
A microcanonical picture emerges for the bosonic mode with equal probability of occupancy for each energy eigenstate $\mu\rangle$ with eigenvalue $\mu$ as shown in Fig.\ \ref{equip}(d), and the diagonal density matrix reduces to the form $\rho_{B}(\mu,\mu) = 1/(N_{max} + 1)$, where $N_{max}$ denotes the largest eigenvalue at which the basis state of the boson is truncated for numerical calculation.
We also calculate the energy of the bosonic mode $E_{B} = \frac{1}{2} \Omega (\langle \hat{Q}^2 \rangle + \langle \hat{P}^2 \rangle)$ at the steady state for different values of $N_{max}$ exhibiting a linear dependence with $N_{max}$ as seen from Fig.\ \ref{equip}(c). This result confirms the microcanonical distribution of the bosonic mode which leads to the analytically obtained expression of its energy $E_{B}= \frac{1}{2} \Omega N_{max}$.
This allows us to draw the notion of an effective temperature $T_{eff}$ and from the equipartition theorem we see that in this case $T_{eff}$ is directly related to the dimension $N_{max}$ of the Hilbert space of the oscillator and is given by,
\begin{equation}
T_{eff} = \Omega N_{max}/2.
\end{equation}        
This result is important since it implies that the KDM thermalize at infinite temperature in the thermodynamic limit i.e $N_{max} \rightarrow \infty$ which has also been observed for other driven many-body systems.

\begin{figure}[ht]
\centering
\includegraphics[scale=0.165]{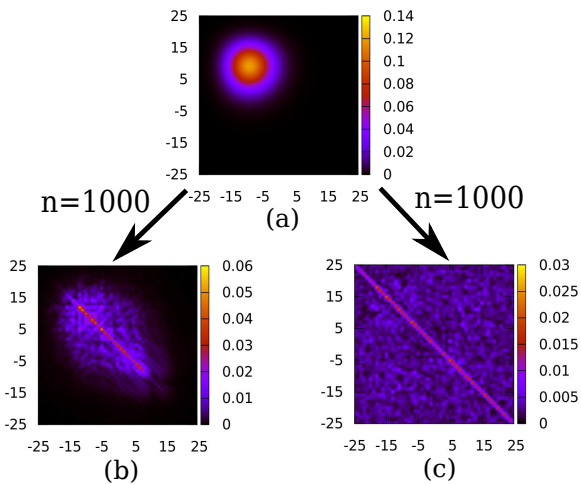}
\caption{Time evolution of the reduced density matrix in spin space, $\rho_S$ has been shown for different kicking strength. (a) The initial density matrix $\vert \psi(0)\rangle \langle \psi(0)\vert$ has been constructed for $S=25$ and $N_{max}=40$ using the initial values $q = 0.11$, $p = 0.14$, $\theta = 1.2$, $\phi = 0.5$. The time evolved reduced spin density matrix has been shown for $\lambda _0=0.4$ in (b) and $\lambda _0=5.0$ in (c) after the time evolution for $n=1000$.}
\label{den_mat}
\end{figure}

Next, we focus on the emergence of thermodynamic distribution of the spin subsystem, which is coupled to the bosonic bath. Dynamical evolution of an initially prepared coherent state reveals that in the chaotic regime (with large $\lambda_{0}$) the reduced density matrix $\rho_{S}$ becomes diagonal in the basis of $\hat{S}_{z}$; on the contrary, non vanishing off-diagonal elements of $\rho_{S}$ implies absence of thermalization for small coupling strength. Clear manifestation of thermalization in the pattern of the density matrix $\rho_{S}$ in two different dynamical regimes is depicted in Fig.\ \ref{den_mat}.  
Since the spin part of DM is entangled with the bosonic bath, it is natural to calculate the entanglement entropy $S_{en} = -Tr\hat{\rho_{S}}\ln\hat{\rho_{S}}$ to quantify the degree entanglement and disorder generated in the spin sub-system.

\begin{figure}[ht]
\centering
\includegraphics[scale=0.5]{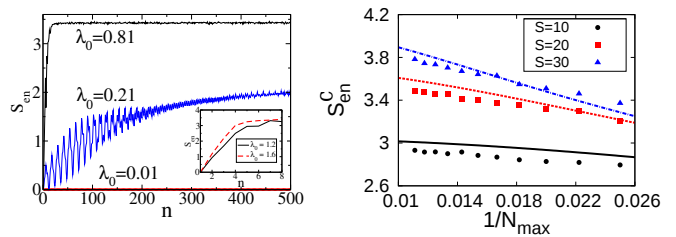}
\caption{(a) Entanglement entropy of the spin space $S_{en}$ as a function of the time step n is shown for different values of $\lambda _0$ with $S=25$ and $N_{max}=50$. (b) The saturated value of $S_{en}$ is plotted as a function of $1/N_{max}$ for $\lambda _0=5.0$; different point types correspond to different values of S and the corresponding lines are obtained from the expression of $S_{en}^{c}$ given in Eq.\ \ref{Entr_can}. Other parameters chosen for this plot are same as in Fig.\ \ref{equip}.}
\label{entn_entropy}
\end{figure}

The time evolution of $S_{en}$ for different coupling strength $\lambda_{0}$ is shown in Fig.\ \ref{entn_entropy}(a). 
Smallness of the entanglement entropy $S_{en}$ in the weak coupling regime signifies the validity of approximating the full wavefunction as product wavefunctions of the sub-systems. For large values of $\lambda_{0}$, we observe that $S_{en}$ increases linearly with time and finally saturates to a steady value (see Fig.\ \ref{entn_entropy}). The linear growth of $S_{en}$ and its steady value in long time dynamics are important signatures of thermalization in chaotic regime\cite{Santos_rev,MRigol_rev}.   
Interestingly, we notice the steady state value of the entropy $S_{en}$ of the spin with fixed magnitude $S$ increases with the dimensionality $N_{max}$ of the bosonic subspace and finally approaches to its maximum value $\log(2S + 1)$ corresponding to the microcanonical ensemble in the thermodynamic limit $N_{max} \rightarrow \infty$.  
This variation of $S_{en}$ with $N_{max}$ can be explained from the fact that the diagonal density matrix $\rho_{S}$ follows canonical distribution with the same effective temperature $T_{eff} = \Omega N_{max}/2$ set by the bosonic bath. Hence the analytical expression of entropy is given by,
\begin{eqnarray}
S_{en}^{c} &=& \ln \left(\frac{\sinh \beta \left(S+\frac{1}{2}\right)}{\sinh \frac{\beta}{2}}\right) + \beta \left[ \frac{1}{2}\coth \frac{\beta}{2} \right.\nonumber \\
&& \left. - \left(S+\frac{1}{2}\right)\coth \beta \left(S+\frac{1}{2}\right) \right]
\label{Entr_can}
\end{eqnarray} 
where $\beta = 1/T_{eff}$. 
In the thermodynamic limit of the bath, i.e $N_{max} \rightarrow \infty$ (equivalently $\beta \rightarrow 0$), the canonical distribution reduces to the microcanonical one; hence the entropy $S_{en}$ converges to its maximum value $\log(2S + 1)$.
From Fig.\ \ref{entn_entropy}(b), we notice that the saturated values of the entropy $S_{en}$ and its variation with $N_{max}$ are well represented by the analytical expression of $S_{en}$ obtained from canonical distribution. 
However for large $N_{max}$ there is no appreciable difference between canonical and microcanonical distribution of $\rho_{S}$, it is possible to differentiate between two ensembles from the deviation of the entropy from its maximum value.
It is important to note that the canonical description of the spin sub-system is valid as long as its dimensionality is much smaller compared to that of bosonic bath, i.e $S<<N_{max}$.

\section{Conclusion}
\label{Conclu}

In this work we present a detailed analysis of quantum and classical dynamics of kicked Dicke model.
It is possible to characterize the dynamics in terms of spectral properties of the Floquet operator.
In the regime of strong kicking strength, the classical system is chaotic and the corresponding quantum system thermalizes to an equilibrium state.

First we have constructed the classical Hamiltonian map from the Heisenberg equation of motion of the corresponding observables 
in an appropriately chosen limit of large spin ($S\rightarrow \infty$) . 
We obtain the fixed points of the classical map for increasing coupling strength which shows an interesting pattern and perform their stability analysis. A critical coupling strength is obtained analytically above which the trivial fixed point becomes unstable and non trivial fixed points appear in pairs; region of stability of these new fixed points decreases with increasing coupling strength. 
By increasing the strength of kicking the phase space trajectories change from regular periodic motion to chaotic dynamics. 

Stroboscopic quantum dynamics is studied using the Floquet operator and a comparison with classical dynamics is presented. 
The average values of the observables obtained from the quantum dynamics are in good agreement with the corresponding quantities calculated from classical map for small values of kicking strength but deviates significantly for increasing coupling strength, particularly in the chaotic regime.
From the time evolution of an initially prepared coherent state, the classicality and coherence of the dynamics is quantified 
by introducing coherence factor, which interestingly in the chaotic regime rapidly decays with a time scale 
in resemblance to the Lyapunov exponent. 
This particular observation is also important for other quantum systems since it raises a question on the validity of semiclassical dynamics in terms of coherent states particularly in the chaotic regime.

We have shown that the phase space properties of the classical system
can be equivalently captured by the spectral properties of quasienergies of the Floquet operator.
For small $\lambda _0$ the phase space dynamics is regular and 
the level spacing distribution of the eigenphases follows the Poisson distribution,
whereas in the regime of strong chaos (for large $\lambda _0$) the corresponding level statistics follows the Wigner distribution. 
As the coupling strength is varied there is a crossover from Poisson to Wigner distribution 
that can be further quantified by computing the average ratio between consecutive level spacing 
$\langle r \rangle$ which interpolates between two limiting values 0.386 (Poisson) and 0.529 (COE). 
The chaotic phase space dynamics is further manifested by the delocalization of the eigenvectors of the Floquet operator; 
the degree of delocalization is quantified by computing the IPR. 
It is noted that in the strong chaotic regime, 
the eigenbasis distribution of the structural entropy becomes highly peaked around $\sim 0.3646$ which is in agreement with the prediction of the random matrix theory.   
This feature suggests that in the regime of strong chaos
all eigenmodes of the Floquet operators have equal contribution.

Finally, motivated by the fact that the system exhibits chaotic phase space dynamics for large coupling strength and the corresponding spectral statistics follows the results of RMT, we have explored whether the chaotic dynamics leads the system to thermalize over time. 
Since in the Dicke model the two subsystems, the spin sector and the bosonic part, are entangled it is very relevant to compute the entanglement entropy for the subsystems. 
For weak kicking strength the entanglement entropy for the spin sector is very small, 
justifying the product wavefunction in the coherent state representation. 
In the strong coupling regime, the 
entanglement entropy increases linearly with time and converges to a steady-state value; 
this linear growth of the entanglement entropy is an indication that the spin sector thermalizes for large kicking strength in the course of time evolution \cite{MRigol_rev,Santos_rev}. 
We have further calculated the reduced density matrices($\rho _{B(S)}$) of the spin subspace and the bosonic mode; 
for large kicking strength the time evolved density matrices turns out to be completely diagonal in the basis of corresponding Hamiltonian.
This is in contrast to the structure of $\rho _B(S)$ in the weak coupling regime
where the off-diagonal elements have significant values even after the long time evolution.
This again strongly indicates that thermalization occurs in the the spin sector for large kicking strength.

Since in the chaotic regime the two subspaces are strongly entangled, it is also important to explore the bosonic subsystem. 
Here also we find that for large kicking strength the reduced density matrix in the bosonic sector, 
in the long time limit, turns out to be diagonal with each element equally weighted by $\sim 1/(N_{max}+1)$,
Furthermore in the the strong coupling regime, $\langle \hat{P}^2 \rangle \simeq \langle \hat{Q}^2 \rangle$ 
exhibiting the equipartition of energy, and the total energy in the bosonic subsystem $E_{OS} \sim N_{max}$. 
These results lead to the conclusion that for large coupling strength an underlying microcanonical 
ensemble picture is emerging with equal probability of occupation in each number state.
In the microcanonical ensemble, it is meaningful to introduce the notion of temperature by $T \sim E_{OS}$,
which can be related to the dimensionality of the bosonic subspace i.e. $T \sim N_{max}$.
In the thermodynamic limit, as the dimensionality $N_{max} \rightarrow \infty$,
the effective temperature of the bosonic bath approaches inifinity \cite{Rigol_drive,Nandkishore}.
Now the signature of canonical distribution in the spin sector, considered to be in contact with a thermal bosonic bath, is captured from the variation of the entanglement entropy $S_{en}$ with $N_{max}$.
In the limit of infinite temperature the canonical distribution reduces to the microcanonical distribution 
and the entropy approaches the well known value $\log(2S + 1)$.
This is remarkable in a sense that it illustrates that the chaotic dynamics of a kicked Dicke model 
has an equivalent and simple  statistical mechanical description of canonical distribution 
in a smaller subsystem(spin) in the presence of a microcanonical bath.

The analysis presented in this work leads us to conclude that such a simple quantum system like the kicked Dicke model with a few collective degrees of freedom can {\it indeed} thermalize as a result of chaotic dynamics of its classical counterpart. In recent experiments on cold atoms the Dicke model have been realized by coupling the BEC with the optical cavity mode\cite{Dicke_exp}. In the similar lines, kicking and periodic drive can also be experimentally generated by using the pulsed cavity mode or by modulating the trap frequency of the condensate. The scenario of thermalization shown in this work can be possibly tested from the statistics of the photon mode.

\end{document}